\documentclass[12pt,draftclsnofoot,onecolumn,peerreview]{IEEEtran}
\usepackage{ifpdf}

\ifCLASSOPTIONcompsoc
  \usepackage[nocompress]{cite}
\else
  \usepackage{cite}
\fi

\ifCLASSINFOpdf
  \usepackage[pdftex]{graphicx}
  \graphicspath{{./figures}{./}}
  \DeclareGraphicsExtensions{.pdf,.jpeg,.png}
\else
\fi

\usepackage{multirow}

\begin{document}
%
\title{What Can Wireless Cellular Technologies Do about the Upcoming Smart Metering Traffic?}

\author{\IEEEauthorblockN{Jimmy J. Nielsen, Germ\'an C. Madue\~no, Nuno K. Pratas, Ren\'e  B. S{\o}rensen, \v Cedomir Stefanovi\'c, Petar Popovski} \\
\IEEEauthorblockA{Department of Electronic Systems, Aalborg University, Denmark\\
Email: \{jjn,gco,nup,cs,petarp\}@es.aau.dk}%
}

\maketitle

\begin{abstract}

The introduction of smart electricity meters with cellular radio interface puts an additional load on the wireless cellular networks. Currently, these meters are designed for low duty cycle billing and occasional system check, which generates a low-rate sporadic traffic. 
As the number of distributed energy resources increases, the household power will become more variable and thus unpredictable from the viewpoint of the Distribution System Operator (DSO). It is therefore expected, in the near future, to have an increased number of Wide Area Measurement System (WAMS) devices with Phasor Measurement Unit (PMU)-like capabilities in the distribution grid, thus allowing the utilities to monitor the low voltage grid quality while providing information required for tighter grid control.
From a communication standpoint, the traffic profile will change drastically towards higher data volumes and higher rates per device. In this paper, we characterize the current traffic generated by smart electricity meters and supplement it with the potential traffic requirements brought by introducing enhanced Smart Meters, i.e., meters with PMU-like capabilities. Our study shows how GSM/GPRS and LTE cellular system performance behaves with the current and next generation smart meters traffic, where it is clearly seen that the PMU data will seriously challenge these wireless systems. We conclude by highlighting the possible solutions for upgrading the cellular standards, in order to cope with the upcoming smart metering traffic.
\end{abstract}

\IEEEpeerreviewmaketitle

\section{Introduction}
\label{sec:introduction}

Smart power grids represent an important group of devices and applications within the umbrella of internet of things (IoT). Especially, the large number of network-connected smart electricity meters that already are or will be located in all households and commercial/industrial locations are representative examples of IoT devices. At present, smart electricity meters are primarily used by electricity providers only for availability monitoring and billing.
However, with the increasing number of distributed energy resources (DERs), such as wind turbines, solar panels, and electric vehicles, strong and sometimes unpredictable variations in the power quality are introduced, prompting for an increased need of monitoring and control.
Specifically, distribution system operators (DSOs) need to be able to observe the circumstances in the low voltage (LV) power grid by introducing more frequently-sampled measurement points. Such wide area measurement systems (WAMS) exist already in the transmission grid, whereas in the distribution grid the DSOs rely mainly on open loop control beyond the substation level, i.e., without real-time feedback from consumers. As the number of DERs increases, this control loop must be closed by providing the feedback from measurements in the LV grid, enabling the state estimation and prediction of the grid behavior, and ultimately ensuring stable operation \cite{sexauer2013phasor}.
It is expected that in the future LV grid, in addition to the traditional smart meter (SM), which is so far primarily used for billing purposes with hourly or daily reports, another more advanced monitoring node will be needed, here referred to as Enhanced Smart Meter (eSM). The eSM is largely similar to a Wide Area Measurement System (WAMS) node, as it integrates Phasor Measurement Unit (PMU)-like capabilities; in other words, the eSM measures power quality parameters (such as power phasors) more frequently and in more details compared to SMs~\cite{ETSI+TR102.935.2012}.
While it is generally expected that not all smart meter locations need to be equipped with eSM devices, the fraction of eSMs needed in the distribution grid to achieve satisfactory state estimation is still an open research question \cite{huang2012state}.

Today, SM devices are typically connected to the DSO backend system using either: 1) a concentrator that gathers the data from the SMs in its neighborhood, via local Wi-Fi or PLC connections, and then relays it via cellular or a wired connection to the DSO backend; or 2) direct connections from each SM through the cellular network to the DSO backend [4].
While the concentrator based approaches reduce the load on the access networks by aggregating data locally, they are not suited for real-time monitoring from eSMs. The reason for this in PLC is the limited bandwidth, reliability and the delays related to the daisy-chain topology. Wi-Fi is challenged due to the issues of the shared spectrum and uncontrollable interference.
Therefore, we assume that SMs and eSMs are equipped with cellular interfaces, so as to eliminate the potential delays, ease deployment and reduce maintenance costs associated with the network connectivity.

The traffic profile generated by smart meters falls into the category of Machine-to-Machine (M2M) traffic.
A main characteristic of M2M traffic is that it consists of transmissions of small amounts of data from a very high number of devices, differing significantly from the bursty and high data rate traffic patterns in human-oriented services, and instead requiring network reliability and availability.
Further, M2M traffic is more demanding in the uplink and less focused on downlink performance, as typical use cases encompass monitoring and control functions.

With LTE gaining an increasing market share, it is expected that within a number of years one of the existing 2G or 3G systems will be taken out of service in order to re-harvest the spectrum to use for newer technologies. 
Current reports on the active M2M cellular devices indicate that $64~\%$ of them are GSM/GPRS-only, $25~\%$ both 3G and GSM/GPRS compatible, $10\%$ 3G-only, and only 1~\% is LTE capable \cite{ericsson2014interim}.
It is clear that GSM/GPRS (hereafter denoted GPRS) dominates the M2M industry, therefore in this paper we analyze how well this technology can support the connectivity demands of SM and eSM devices.
Given the promise of LTE, we also investigate its potential use as the access network for eSM devices.

Specifically, in this paper we have the following four contributions: 1) extraction and classification of smart meter traffic models from relevant specifications, as well as predicted future traffic growth; 2) comprehensive simulation model of radio access systems that includes all phases in the access, in contrast to \cite{hagerling2014coverage} and the NIST \emph{PAP2 guidelines for assessing wireless standards for smart grid application v1.0} that use simplified models; 3) quantitative assessment of how many smart meter devices can be supported in cellular systems, comparing the simplified and comprehensive simulation model results; and 4) recommendations for standardization and future roadmap of the radio access technologies.

The rest of this paper is organized as follows.
In Section~\ref{sec:smart_meter_traffic_model} and \ref{sec:wams_traffic_model} we characterize the traffic models of the SM and eSM devices.
In Section~\ref{sec:cellular_systems_performance} we describe the access bottlenecks in the cellular access reservation protocol and provide numerical results that show how the proposed traffic models affect the performance of GPRS and LTE networks.
Then, in Section~\ref{sec:standardization_outlook} we provide guidelines that the future cellular network standards should take in account when designing the network system.
Finally, we wrap-up the paper with the main take-home points.

\section{Smart Meter Traffic Model} 
\label{sec:smart_meter_traffic_model}

In the literature there are different examples of traffic models for traditional smart meters.
Of these, the OpenSG \emph{Smart Grid Networks System Requirements Specification} (described in \cite{hossain2012smart}) from the Utilities Communications Architecture (UCA) user group is the most coherent and detailed network requirement specification, and it has therefore been used in this paper as input for the SM traffic model. The UCA OpenSG is a relevant consortium of 190 companies and the considered smart grid use cases are in line with those studied by other standardization organizations such as ETSI and USEF.
Since there are differences in which use cases and applications are offered by the DSO or electricity retail company and which of those the individual customers are using, a one size fits all traffic model does not exist.
In the following we consider a comprehensive configuration where all use cases that involve communication from the smart meters to the core network will be in operation and note that actual deployments with different configurations may lead to different results. 
For calculating the message frequency in the uplink SM traffic model the event occurrence frequencies listed in Table~\ref{tab:event_assumptions} have been used.
Besides the values listed in Table~\ref{tab:event_assumptions} we assume that a commercial/industrial SM sends a 2400 bytes meter reading packet every hour, whereas a residential SM sends a 1200 bytes report every 4 hours.
\begin{table}[tbh]
\centering
\begin{tabular}{l|l} 
\textbf{Event} & \textbf{Frequency [events per meter]} \\ \hline
On-demand meter read requests & 25/1000 per day \\
Meter capped energy mode request & 5 per year \\
DR load management request to HAN devices & 15/1000 per day \\
HAN device join/unjoin & 5 per year \\
Real-time price (RTP) update & 96 per day \\
Metrology firmware update  & 4 per year \\
Metrology program update  & 4 per year \\
NIC firmware update & 4 per year \\
NIC program update & 4 per year \\
\end{tabular}
\caption{Assumptions for deriving traffic model.}
\label{tab:event_assumptions}
\end{table}

\begin{figure}[tbh]
  \begin{center}
    \includegraphics[scale=0.75]{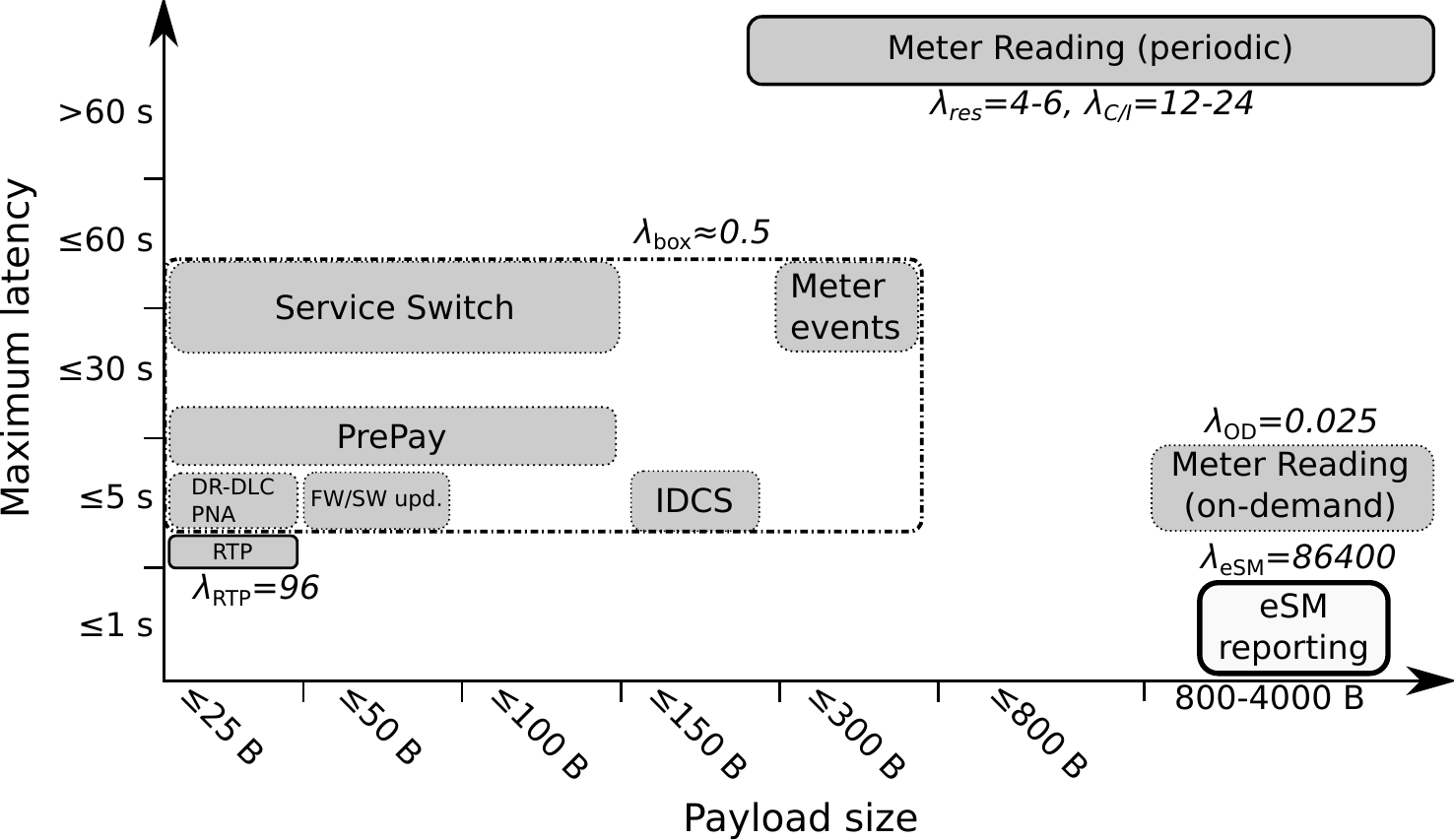}
  \end{center}
  \caption{Classification of OpenSmartGrid traffic originating from an SM. $\lambda$-values show the number of generated messages per day per device. Use case short names: Demand Response - Direct Load Control (DR-DLC), Premise Network Administration (PNA), Firmware and Software updates (FW/SW upd.), Real-Time Price (RTP), Islanded Distributed Customer Storage (IDCS).}
  \label{fig:categorization_PL-latency}
\end{figure}

The SM uplink traffic model, resulting from the above assumptions, is presented in Fig.~\ref{fig:categorization_PL-latency}. The gray boxes represent the different use cases and the boxes span the latency and payload size requirements of the corresponding messages. The white box represents the eSM traffic (defined in Section \ref{sec:wams_traffic_model}).
Nearly all use cases have reliability requirements of $98\%$, with the exceptions being two alarm messages in the IDCS use case requiring $99\%$, and the periodic meter reading, which has time-dependent reliability requirements ranging from $98\%$ to $99.5\%$. 
In relation to the figure, Table~\ref{tab:ul-dl_ratio} shows the average estimated uplink/downlink bandwidth for each use case.

The $\lambda$-values in Fig.~\ref{fig:categorization_PL-latency} shows the number of generated messages per day per SM. The use cases grouped in the dash-dotted box transmit very infrequently with a combined rate of only $\sim0.5$ messages per day. Further, they are relatively similar in terms of latency and payload size. In addition to this group, two other OpenSG use cases from the figure stand out, namely the real-time pricing (RTP) that causes 96 messages per day and the periodic meter reading on the top right. For periodic meter readings, a commercial/industrial (C/I) SM sends reports more often than a residential SM.
Notice for the eSM reporting that, in addition to the stricter latency requirement of $\leq 1$ sec, the number of generated messages per day is many orders of magnitude higher than any of the SM use cases.

\begin{table}[!htb]
  \centering
  \begin{tabular}{|lr|r||r|r|r|r|r|} \cline{3-8}
  \multicolumn{2}{c|}{} & \textbf{downlink} & \multicolumn{5}{c|}{\textbf{uplink}} \\ \hline
   \textbf{Use case} & \textbf{RI} & \textbf{default} & \textbf{default} & \textbf{5 min} & \textbf{1 min} & \textbf{30 sec} & \textbf{15 sec} \\ \hline
  \multicolumn{2}{|l|}{Meter Reading} & 1.25 & 11K & 95K & 475K & 950K & 1.9M \\
  \multicolumn{2}{|l|}{Service Switch} & 3 & 6 & 6 & 6 & 6 & 6 \\
  \multicolumn{2}{|l|}{PrePay} & 3.5 & 8 & 8 & 8 & 8 & 8 \\
  \multicolumn{2}{|l|}{Meter Events} & 0 & 50 & 50 & 50 & 50 & 50 \\
  \multicolumn{2}{|l|}{Islanded Distr. Cust. Storage} & 2 & 5 & 5 & 5 & 5 & 5 \\
  \multicolumn{2}{|l|}{DR-DLC} & 400 & 0.5 & 0.5 & 0.5 & 0.5 & 0.5 \\ 
  \multicolumn{2}{|l|}{Premise Network Admin} & 1 & 1 & 1 & 1 & 1 & 1 \\
  \multicolumn{2}{|l|}{Price} & 10K & 2.4K & 2.4K & 2.4K & 2.4K & 2.4K \\
  \multicolumn{2}{|l|}{Firmware / Program Update} & 30K & 5 & 5 & 5 & 5 & 5 \\ \hline
  \multicolumn{2}{|l|}{Total} & 40.4K & 13.4K & 97K & 477K & 952K & 1.9M \\ \hline
  \end{tabular}
  \vspace{0.2cm} 
   \caption{Average downlink/uplink raw data rate as [bytes/meter/day] for the considered use cases. Default value of RI is 4 hours for residential and 1 hour for commercial/industrial SMs.}
    \label{tab:ul-dl_ratio}
\end{table}

Table~\ref{tab:ul-dl_ratio} shows that the raw data rate requirements of SMs with default reporting interval (RI) are quite  modest, with an average uplink data rate of appr. 13.4KB per day per SM and an average downlink data rate of appr. 40.4KB per day per SM. While the total downlink data rate is actually higher than the uplink, it is constituted primarily of software updates, which are large low-priority data transfers that occur infrequently during the night, where it does not interfere with the day-to-day operation of the smart grid.
Given the modest traffic requirements, it is expected that GPRS networks, that are deployed ubiquitously and offer a reliable coverage, but are gradually becoming less suitable for human-oriented traffic, can easily satisfy the default SM traffic requirements.

Further, an option to increase observability in the power grid is to reduce the meter reading reporting interval.
We investigate how capable the current cellular systems are to support this in Section~\ref{sec:cellular_systems_performance}, when the report packet sizes are respectively 300 bytes and 600 bytes for residential and commercial/industrial and reporting intervals range from 5 min, 1 min, 30 sec, to 15 sec.
As shown in Table~\ref{tab:ul-dl_ratio}, in case of these reduced RIs, the uplink data rate requirements become much larger than in the downlink.


\section{Enhanced Smart Meter Traffic Model} 
\label{sec:wams_traffic_model}

The eSM is a PMU-like device for the distribution grid, which is able to measure voltage and current phasors. However, it has less strict real-time requirements than transmission grid PMUs, since it is used to increase observability rather than for protection purposes. 
Being deployed not only in DSO substations, but also in and close to prosumer homes, the eSM reports measurements through cellular networks, since this allows a mobile network operator to prioritize and dedicate resources to eSM traffic, thus achieving QoS, which may not be possible 3rd party consumer-grade wired Internet connections.
Phasor measurements can be used on different time scales, ranging from a few milliseconds (e.g., for protective relays) up to several seconds (e.g., real-time monitoring and state estimation) \cite{sexauer2013phasor}.
The eSMs are intended to improve observability and enable state estimation and real-time control \cite{sexauer2013phasor}, with the suggested lower bound of 1 second for the reporting interval~\cite{adamiak2005synchrophasors}.
Since the eSMs features and requirements are not yet standardized, the eSM traffic model considered in this study is based on the requirements of the transmission grid PMU and WAMS related standards, IEEE 1588, IEEE C37.118 and IEC 68150.
Specifically, we assume that every second an eSM sends a measurement report that consist of concatenated PMU measurements (50~Hz sample rate) from the preceding 1 second measurement interval. The samples are, as specified in PMU standards IEEE 1588 and C37.118, timestamped using GPS time precision. 
Assuming that the floating point PMU frame format from IEEE 1588 is used and that each sample covers 6 phasors, 1 analog value and 1 digital value, each PMU frame accounts to 76 bytes. Adding UDP header (8 bytes) and IPv6 header (40 bytes) to each report of 50 PMU samples, an eSM packet is 3848 bytes, and a bit rate of 30.8 kbit/s. 
Since it may be an exaggeration to send all 50 PMU samples per measurement interval, we also consider in our performance analysis the case of eSM reduced report sizes.


\section{Cellular Systems Performance}
\label{sec:cellular_systems_performance}

From the communications perspective, it is important to investigate which cellular technologies can support the current billing-only smart meter use cases, but also the use cases/services that go beyond the current ones. 
In \cite{hagerling2014coverage} and NIST \emph{PAP2 guidelines for assessing wireless standards for smart grid application v1.0} performance analyses were carried out to determine the number of smart grid devices supported by different wireless technologies, however, they only evaluated the data capacity of the systems and neglected to account for the bottlenecks in the access reservation protocol used in cellular systems. As it is shown in \cite{7007669}, the access reservation bottlenecks are particularly prone to exposure with M2M traffic such as smart grid traffic, meaning that a pure data capacity based analysis may lead to overly optimistic results. Therefore, our analysis will include all aspects of the access reservation procedure and compare those results to a data capacity only analysis. For the analysis we will consider the traffic patterns for SM and eSM devices described in Sections \ref{sec:smart_meter_traffic_model} and \ref{sec:wams_traffic_model}.
From those traffic models it is clear that the communication requirements of these two device types are orders of magnitudes apart in terms of message frequency and bandwidth, meaning that for eSM deployment a more capable technology than GPRS is needed. With its integrated PMU unit, the eSM is already a more complex and expensive device than the SM, and since fewer eSMs than SM will be needed, a higher unit price can be better tolerated, and thus we will assume that the eSM uses LTE.


\subsection{Access Reservation Protocol Operation and Limitations} 
\label{sub:protocol_limitations}

In cellular networks, a device with no active connection to the network first has to establish one, in order to perform a data transmission.
This is accomplished via an access reservation protocol, which in general consists of three main stages: random access, granting access and data access.
In the first stage, the cellular devices perform a random access request in one of the random access opportunities (RAOs). In the second stage, the base station grants access to the network if: (i) the random access request is received without error by the base station; (ii) no other device has transmitted in the same RAO (i.e., collision free); and (iii) there are data resources available to the device.
Otherwise, the access reservation procedure must be restarted and the device will transmit a new random access request until it is granted by the base station or until the maximum number of retransmissions is reached and the request fails.
In GPRS there are 217 RAOs/s per carrier while in LTE there are 10.8k RAOs/s\footnote{Assuming the contention resources occur every 5 ms, each with 54 contention preambles available.}.
On the other hand, only 32 grants/s and 3k grants/s are offered in GPRS and LTE, respectively~\cite{6820749,7007669}.
Therefore, when the random access stage is heavily loaded, the grant stage becomes a decisive limitation in cellular networks.
Furthermore, in GPRS and LTE the data stage is not only limited by the amount of the actual data resources, but also by the amount of the uplink identifiers used to coordinate transmissions from active devices, which limit the amount of simultaneously active M2M communication links.

\subsection{Outage Performance Evaluation} 
\label{sub:cellular_system_performance_results}

To evaluate the performance of the cellular access, we used the outage rate, i.e., the probability of a device failing to deliver a report before the report deadline expires, while accounting for the access reservation protocol.
This can be regarded as a measure of the cellular access reliability, which is the paramount performance indicator for the wide-area distribution supervision and control applications \cite{goel2013ieee}. 
To evaluate the outage we used event-driven simulators developed in MATLAB, which cover the complete access reservation protocol, as defined in 3GPP Release 12.
Particularly, the GPRS simulator considers the amount of available access granted messages in the access granted channel (AGCH), with a typical configuration of 28 AGCH/s \cite{7007669}, the limited number of the identifiers used to coordinate the uplink transmissions, i.e., the uplink stage flag (USF), and the amount of data resources available.
The LTE simulator considers the restricted amount of access grant messages (RAR messages) due to the physical downlink control channel (PDCCH) limitations~\cite{3GPPTR23887,6781592}. 
Finally, both in the GPRS and the LTE simulator, the data resources are shared with the signaling required in the access reservation procedure and the actual data transmissions.
\begin{figure}[tb]
  \begin{center}
    \includegraphics[width=\linewidth]{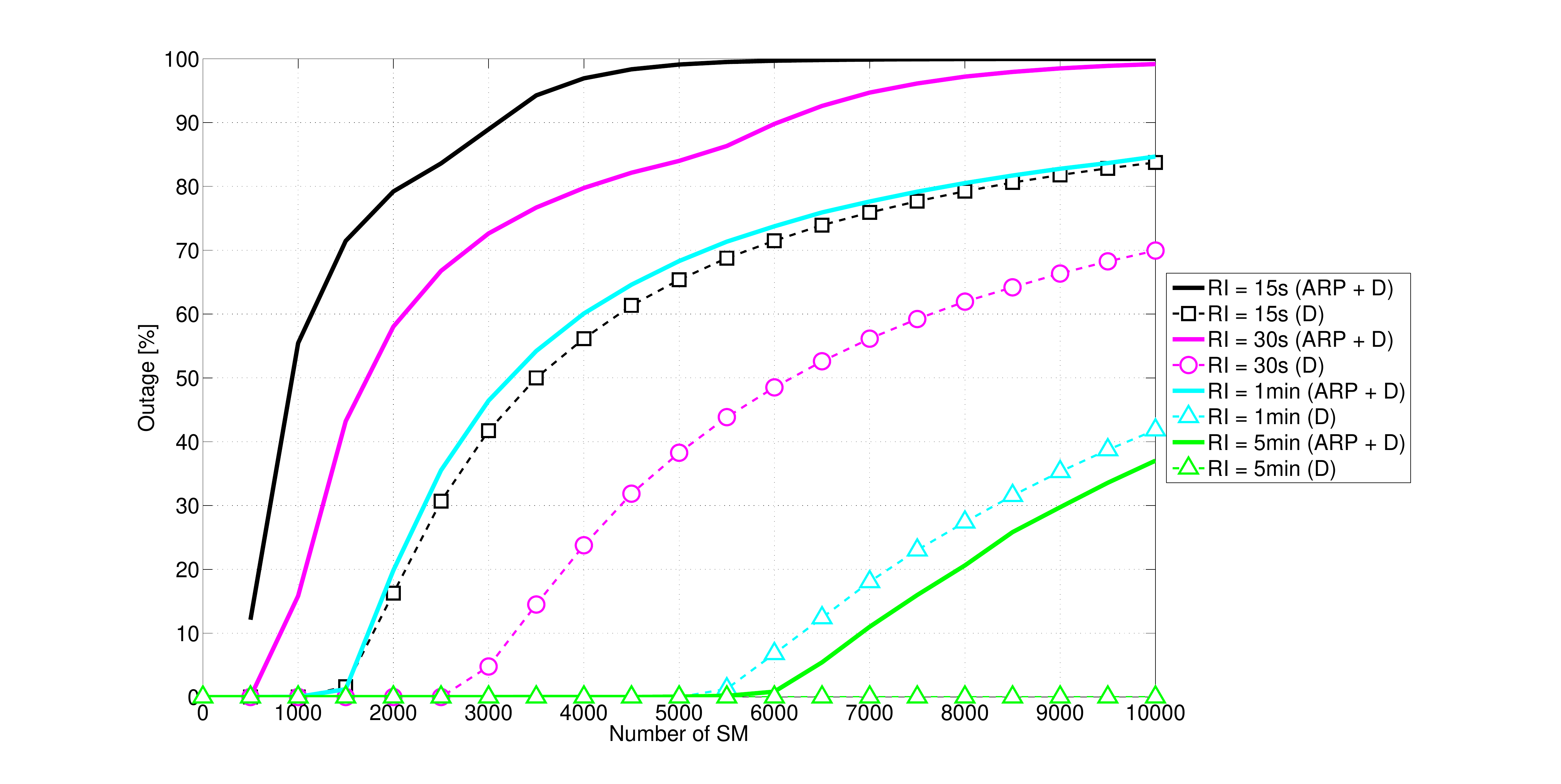}
  \end{center}
  \caption{GPRS outage evaluation for increasing number of SM with different report interval values and RS = 300 bytes for residential and RS = 600 bytes for commercial/industrial, where ARP+D denotes the access reservation protocol plus data phase, while D denotes only data phase.}
  \label{fig:GPRSPlot}
\end{figure}
The evaluation scenario is set in a single cell with 1000~m radius, which includes 4500 smart meters~\cite{7007669}, from which $90$\% correspond to residential customers and the remaining $10$\% to commercial/industrial customers.
In the case of GPRS, we consider a single carrier corresponding to a 200~KHz system.
The considered LTE bandwidth is $1.4$~MHz (6~PRBs), in line with the reduced capabilities for LTE devices~\cite{3GPPRelease13Overview}.
In addition, the control channel and data channel probability of error, are respectively $10^{-2}$ and $10^{-1}$~\cite{6820749}.
In both systems, we assume the devices always transmit with the highest modulation scheme available, in order to focus the evaluation on the performance of the access reservation protocol.
In these conditions, we observed that the SM traffic, provided in Section~\ref{sec:smart_meter_traffic_model}, is supported by both GPRS and LTE with near 0\% outage, as the total number of messages per hour from each SM only amounts to approximately 125.
\begin{figure}[tb]
  \begin{center}
    \includegraphics[width=\linewidth]{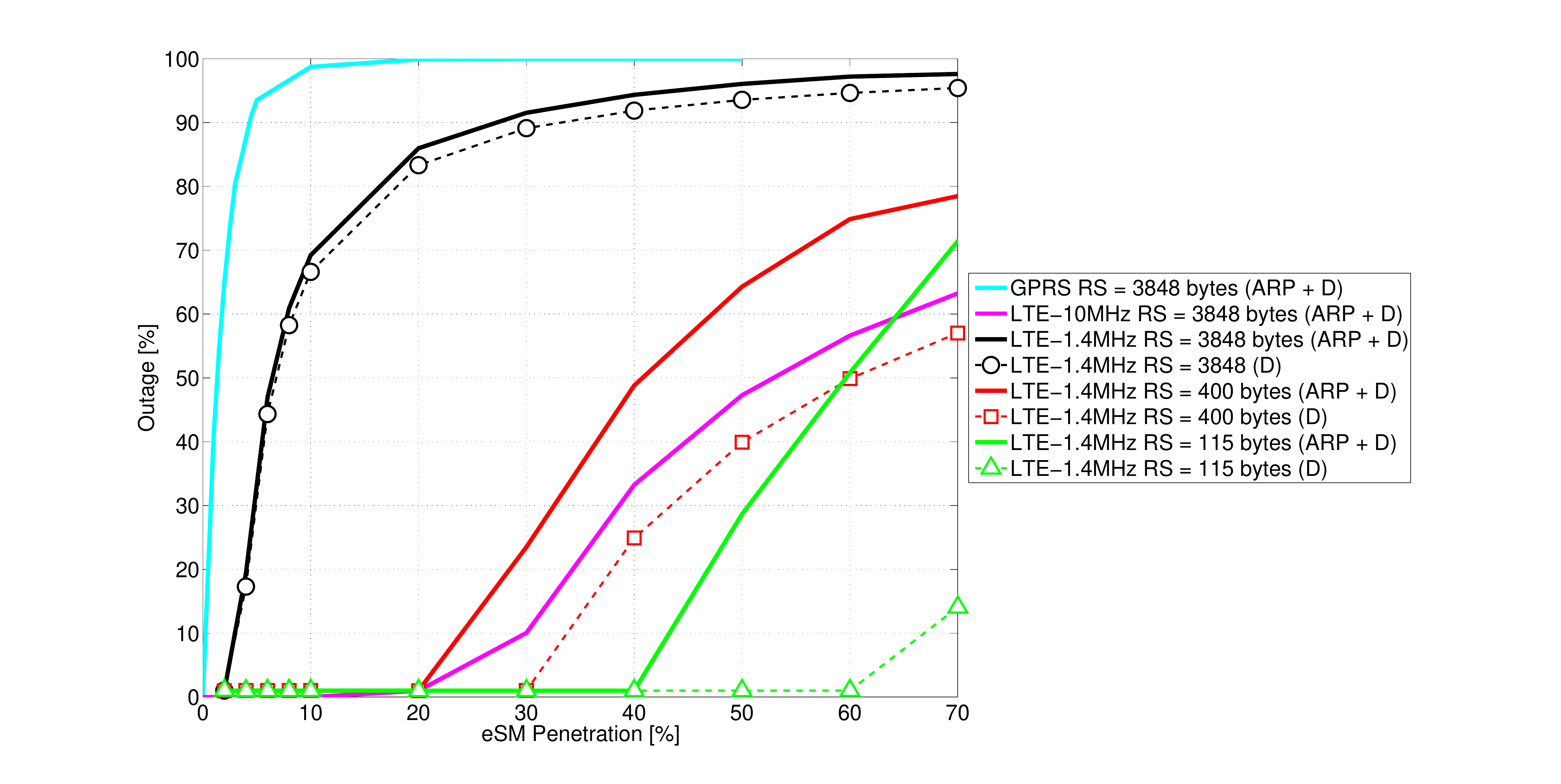}
  \end{center}
  \caption{LTE and GPRS outage evaluation for increasing penetration of eSMs, where ARP+D denotes the access reservation protocol plus data phase, while D denotes only data phase.}
  \label{fig:LTE_and_GPRSPlot}
\end{figure}
We start by considering for GPRS the scenario of reducing SM Report Intervals (RI), as described in Section~\ref{sec:smart_meter_traffic_model}.
Fig.~\ref{fig:GPRSPlot} depicts the outage probability for increasing number of SMs and different RIs. Taking as reference a cell population of $4500$ SMs, we can see that for RI $>$ 5 min, GPRS can provide a significant increase on the distribution network observability from hourly intervals to every 5 minutes.
For smaller report intervals to be supported in GPRS, then the options are either to reduce the cell size and/or increase the number of carriers.  

We proceed by considering in Fig.~\ref{fig:LTE_and_GPRSPlot} the cellular network outage as a function of the eSM penetration, i.e., of how many eSMs are deployed per every 100 smart meter locations. 
As described in Section~\ref{sec:wams_traffic_model}, each eSM report contains 50 samples of the power phasors measured since the last report with an expected payload of $3848$~bytes.
Since this large payload has severe implications on the cellular network performance, we also consider the impact of smaller payloads on system performance, which can be motivated by the introduction of pre-processing to extract statistics, data compression and/or reduced number of samples.
Specifically, we consider reduced report sizes (RS) of 3848, 400, and 115~bytes, where the last two values correspond respectively to a payload reduction of approximately 10\% and 3\% of the original payload size.
The outage results for LTE and GPRS are shown in Fig.~\ref{fig:LTE_and_GPRSPlot}.
We note that GPRS is not able to support eSM traffic irrespective of the chosen RS, while LTE for RS of 3848~bytes only supports up to 2\% eSM penetration.
When a 10 MHz bandwidth is completely dedicated in LTE to serve the eSM traffic then it is possible to reach 30\% of penetration with less than 10\% of outage, which means a large amount of resources dedicated to a potentially low profit application.
On the other hand, if we assume lower RS, already at 400~bytes LTE supports up to 20\% of eSMs.
Further, when comparing the results that correspond to the case when only data phase is taken into account with the results obtained by considering the access reservation phase as well, it can be  observed that the access reservation protocol impacts the number of supported eSMs. 
Particularly, the limitations of the access reservation protocol become substantial as the report size decreases and it could shown that this is mainly due to the lack of access grant messages 
required to complete the access reservation procedure.
Note that this effect has been overlooked in the previous works \cite{nist2011pap2,hagerling2014coverage}.

The presented results allow us to conclude that the RS of the eSM nodes must be small to support a high percentage of nodes.
In addition, we emphasize that small data traffic cannot be analyzed only in terms of the system data capacity, but that the bottlenecks of the access reservation protocol itself must be considered, as observed in the gap between the two types of analysis depicted in Fig.~\ref{fig:LTE_and_GPRSPlot}.
We conclude by noting that in practice, when deploying eSMs, due to the required communication reliability, good coverage should be ensured, e.g., by careful selection of the placement location and/or by adding an external antenna if needed.
In the above presented study, it is assumed that all SMs and eSMs are under a cellular coverage.

\section{Standardization Outlook} 
\label{sec:standardization_outlook}

Although the traffic resulting from smart meters can be easily accommodated into current cellular systems, the same is not observed for the traffic generated by the eSM. In the following, we discuss the challenges and possible solutions that need to be tackled by standardization bodies to ensure that the observability of the distribution network can be improved efficiently.

\subsection{Smart Meter} 
\label{sub:higher_layers}

The inclusion of additional phasor measurement units into the distribution grid, so as to increase its observability, is being discussed specifically at the last mile to the customer premises~\cite{huang2012state}.
Currently, it is not yet clear if that will imply the same level of detail (in number of samples and report frequency) as in the transmission grid PMUs, where the reporting is done by SCADA over dedicated wired links.

As discussed in Section~\ref{sec:cellular_systems_performance}, if the eSMs generate the same amount of traffic as transmission grid PMUs, then the cellular networks will require an extensive overhaul to be able to support both eSM and human centric traffic, leading to substantial investment in the cellular infrastructure.
On the other hand, eSMs will most likely be lighter versions of PMUs, both sampling and reporting less frequently. Therefore, if local processing and compression of the monitoring data is allowed and/or the required level of detail lowered, then the amount of generated traffic will be much lower.
Another viable option, as discussed in Section~\ref{sec:cellular_systems_performance}, is to increase the report frequency of current smart meters without introducing local PMU functionality. The generated small packets could then be handled by the network, as long as the bottlenecks at the access protocol level are addressed.

It seems likely that the standardization for the eSM's PMU functionality falls within the scope of the IEEE C37.118 and IEC 68150 standards, since these specify the measurement and communications requirements for traditional PMU units.
Therefore, it is of paramount importance that standardization bodies reach a consensus on the eSM communication requirements allowing the affected stakeholders to take informed actions.


\subsection{Cellular Network} 
\label{sub:lower_layers}

In 3GPP, the standardization body responsible for the cellular air interface and core network functionality, there are two activities that will affect how the traffic from SM and eSM will be handled~\cite{3GPPRelease13Overview}.

We start by noting that, although GRPS is seen as an outdated communication technology~\cite{3GPPRelease13Overview}, there is an ongoing effort to continue to reengineer GPRS to serve M2M applications, in which the SM traffic can be classified. One of the goals of this initiative is to achieve\footnote{Considering the minimum SDU size, i.e. 80 bytes, with 4 seconds latency.} 160 bit/s.
Concurrently, there is a push from the industry (both utilities and vendors) to keep GPRS networks and their associated resources active, while facing the pressure to re-harvest the GPRS spectrum to be used in the next cellular network generation.
A viable solution to keep the GPRS connectivity, is to \emph{virtualize} its air interface into the next generation cellular systems. 

The second effort is to define a low complexity LTE user equipment category with respect to the cellular interface, which supports reduced bandwidth and transmit power while extending coverage operation~\cite{3GPPRelease13Overview}. Specifically, the goal of reduced bandwidth is to specify 1.4~MHz operation within any LTE system bandwidth, allowing operators to multiplex reduced bandwidth MTC devices and regular devices within their existing LTE deployments. In terms of extended coverage the goal is to improve the coverage of delay-tolerant MTC devices by 15 dB, thereby allowing operators to reach MTC devices in poor coverage conditions, such as smart meters located in basements~\cite{3GPPRelease13Overview}.

To further improve the support of the traffic generated by SM and eSM with very low duty cycle and latency requirements in the order of seconds, the inclusion of \emph{periodic reporting and discontinuous transmission} functionality into cellular standards should be considered. In here, the network provides periodic communication resources so that devices can perform their short data transmission. This allows devices to go to sleep and save energy, since they have prior knowledge of when the next transmission time slot can occur. A solution based on this concept has been proposed through the reengineering of the LTE access protocol~\cite{6820749}.

To cope with the eSM traffic demands and increase the network capacity, \emph{localized aggregation of traffic} should be considered.
In this solution the traffic generated by multiple SMs and eSMs in a geographical area could be aggregated, at eSMs or cellular relays, and then trunked to the cellular network~\cite{Rigazzi2015}.
The use of aggregation and relaying would then allow to decrease the contention pressure at the base station, as well as to improve the single link connection, providing connectivity and coverage enhancements to SMs and eSMs with poor propagation conditions. 

Finally, to support \emph{massive asynchronous access of small packet transmissions}, access reservation protocols in cellular systems are just the first step of the asynchronous access to the network. After it has been completed, then the device starts exchanging signaling information via the higher layers with the entities in the core network, which leads to a high signaling overhead and possible air interface and core network congestion. Although there are already efforts to reduce the signaling exchanges with the core network~\cite{3GPPTR23887}, when the payloads are small enough, the facility to perform the data transmission already in the third step of the access reservation protocol should be in place.

\section{Conclusion} 
\label{sec:conclusion}

In this paper we have evaluated two approaches to increase the observability of the network: (1) decreasing the report interval of the meter reading and (2) introduction of enhanced smart meters with phasor measurement units (PMUs).
We provided details on the characteristics of the traffic generated by smart meters and enhanced smart meters and have highlighted the associated challenges in supporting it from a cellular network point of view.
The obtained results show that GPRS can support traditional smart meter traffic, as well as more frequent measurements down to 5 min report intervals.
Further, it is shown that LTE can support distribution grid PMUs, if the report payloads are appropriately dimensioned.
These results can be used as input for both smart meter and cellular system standardization bodies to enable the introduction of current and future smart grid devices into the cellular networks.
The current main open issue is the uncertainty associated with the eSM communication requirements, which will lead to different cellular systems optimizations.


\section*{Acknowledgment}
This work is partially funded by EU, under Grant agreement no. 619437 ''SUNSEED''. 
The SUNSEED project is a joint undertaking of 9 partner institutions and their contributions are fully acknowledged.
The work of  \v C. Stefanovi\'c was supported by the Danish Council for Independent Research, grant no. DFF-4005-00281.

\end{document}